\documentstyle[12pt,epsf]{article}
\def\fo{\hbox{{1}\kern-.25em\hbox{l}}}
\def\fnote#1#2{\begingroup\def\thefootnote{#1}\footnote{#2}\addtocounter
{footnote}{-1}\endgroup}

\renewcommand{\thefootnote}{\fnsymbol{footnote}}

\def\beq{\begin{equation}}
\def\eeq{\end{equation}}
\def\eq{\end{equation}}
\def\to{\rightarrow}

\def\bsg{\ifmmode B\to X_s\gamma\else $B\to X_s\gamma$\fi}
\def\bsll{\ifmmode B\to X_s\ell^+\ell^-\else $B\to X_s\ell^+\ell^-$\fi}
\def\bstt{\ifmmode B\to X_s\tau^+\tau^-\else $B\to X_s\tau^+\tau^-$\fi}
\def\shat{\ifmmode \hat{s}\else $\hat{s}$\fi}

\newcommand{\newc}{\newcommand}

\newc{\lcal}{\int {\cal L}dt}

\newc{\mHpm}{m_{H^\pm}}
\newc{\gsim}{\lower.7ex\hbox{$\;\stackrel{\textstyle>}{\sim}\;$}}
\newc{\lsim}{\lower.7ex\hbox{$\;\stackrel{\textstyle<}{\sim}\;$}}
\newc{\ie}{{\it i.e.}}          
\newc{\etal}{{\it et al.}}
\newc{\eg}{{\it e.g.}}          
\newc{\kev}{\hbox{\rm\,keV}}            
\newc{\mev}{\hbox{\rm\,MeV}}            
\newc{\gev}{\hbox{\rm\,GeV}}            
\newc{\tev}{\hbox{\rm\,TeV}}
\newc{\xpb}{\hbox{\rm\, pb}}
\newc{\xfb}{\hbox{\rm\, fb}}

\newc{\mtop}{m_t}
\newc{\mbot}{m_b}
\newc{\mz}{m_Z}
\newc{\mw}{M_W}
\newc{\alphasmz}{\alpha_s(m_Z^2)}
\newc{\swsq}{\sin^2\theta_W}
\newc{\tw}{\tan\theta_W}
\newc{\cw}{\cos\theta_W}
\newc{\sw}{\sin\theta_W}
\newc{\BR}{\hbox{\rm BR}}
\newc{\zbb}{Z\to b\bar}
\newc{\Gb}{\Gamma (Z\to b\bar b)}
\newc{\Gh}{\Gamma (Z\to \hbox{\rm hadrons})}
\newc{\rbsm}{R_b^\hbox{\rm sm}}
\newc{\rbsusy}{R_b^\hbox{\rm susy}}
\newc{\drb}{\delta R_b}

\newc{\sgn}{\mbox{sgn}}
\newc{\tbeta}{\tan\beta}
\newc{\uL}{{\tilde u_L}}
\newc{\uR}{{\tilde u_R}}
\newc{\cL}{{\tilde c_L}}
\newc{\cR}{{\tilde c_R}}
\newc{\tL}{{\tilde t_L}}
\newc{\tR}{{\tilde t_R}}
\newc{\dL}{{\tilde d_L}}
\newc{\dR}{{\tilde d_R}}
\newc{\sL}{{\tilde s_L}}
\newc{\sR}{{\tilde s_R}}
\newc{\bL}{{\tilde b_L}}
\newc{\bR}{{\tilde b_R}}
\newc{\eL}{{\tilde e_L}}
\newc{\eR}{{\tilde e_R}}
\newc{\mhp}{m_{H^\pm}}
\newc{\mhalf}{m_{1/2}}

\newc{\lR}{\tilde{l}_R}
\newc{\lL}{\tilde{l}_L}
\newc{\nL}{\tilde{\nu}_L}
\newc{\na}{\chi^0_1}
\newc{\nb}{\chi^0_2}
\newc{\nc}{\chi^0_3}
\newc{\nd}{\chi^0_4}
\newc{\ca}{\chi^{\pm}_1}
\newc{\cb}{\chi^{\pm}_2}
\newc{\camp}{\chi^\mp_1}
\newc{\cbmp}{\chi^\mp_1}
\newc{\capos}{\chi^{+}_1}
\newc{\caneg}{\chi^{-}_1}
\newc{\phit}{\phi_t}
\newc{\phib}{\phi_b}
\newc{\phiew}{\phi_{ew}}
\newc{\htz}{h^0_t}
\newc{\hbz}{h^0_b}
\newc{\hewz}{h^0_{ew}}
\newc{\hsmz}{h^0_{sm}}

\def\NPB#1#2#3{Nucl. Phys. B {\bf #1} (19#2) #3}
\def\PLB#1#2#3{Phys. Lett. B {\bf #1} (19#2) #3}

\def\PRD#1#2#3{Phys. Rev. D {\bf #1} (19#2) #3}
\def\PRL#1#2#3{Phys. Rev. Lett. {\bf#1} (19#2) #3}

\def\ZPC#1#2#3{Zeit. f\"ur Physik C {\bf #1} (19#2) #3}

\def\beq{\begin{equation}}
\def\eeq{\end{equation}}
\def\bea{\begin{eqnarray*}}
\def\eea{\end{eqnarray*}}
\def\slashchar#1{\setbox0=\hbox{$#1$}           % set a box for #1
   \dimen0=\wd0                                 % and get its size 
   \setbox1=\hbox{/} \dimen1=\wd1               % get size of /
   \ifdim\dimen0>\dimen1                        % #1 is bigger 
      \rlap{\hbox to \dimen0{\hfil/\hfil}}      % so center / in box 
      #1                                        % and print #1
   \else                                        % / is bigger 
      \rlap{\hbox to \dimen1{\hfil$#1$\hfil}}   % so center #1
      /                                         % and print /
   \fi}                                         %
\catcode`@=11
\long\def\@caption#1[#2]#3{\par\addcontentsline{\csname
  ext@#1\endcsname}{#1}{\protect\numberline{\csname
  the#1\endcsname}{\ignorespaces #2}}\begingroup
    \small
    \@parboxrestore
    \@makecaption{\csname fnum@#1\endcsname}{\ignorespaces #3}\par
  \endgroup}
\catcode`@=12

\def\jfig#1#2#3{
 \begin{figure}
 \centering
 \epsfysize=3.5in
 \hspace*{0in}
 \epsffile{#2}
 \caption{#3}
 \label{#1}
 \end{figure}}
\def\sfig#1#2#3{
 \begin{figure}
 \centering
 \epsfysize=2.5in
 \hspace*{0in}
 \epsffile{#2}
 \caption{#3}
 \label{#1}
 \end{figure}}
\begin{document}

\begin{titlepage}

\begin{flushright}
SLAC-PUB-7374 \\
hep-ph/9612292\\
December 1996
\end{flushright}

%\vspace{.3in}
%\begin{center}
%ROUGH DRAFT, DO NOT DISTRIBUTE
%\end{center}

\huge
%\vspace{0.05in}
%\renewcommand{\thefootnote}{\fnsymbol{footnote}}
\begin{center}
The electroweak symmetry breaking \\ 
Higgs boson in models with  \\
top-quark condensation
\end{center}
%\vspace{0.2in}

\large

\vspace{.15in}
\begin{center}

James D. Wells \\

\vspace{.1in}
{\it Stanford Linear Accelerator Center \\
Stanford University, Stanford, CA 94309\fnote{\dagger}{Work 
supported by the Department of Energy
under contract DE-AC03-76SF00515.}\\}

\end{center}
 
%\vspace{0.2in}
 
\vspace{0.15in}
 
\normalsize

\begin{abstract}

The top quark may get its large mass not from a fundamental
scalar but a Nambu-Jona-Lasinio mechanism involving a strongly
coupled gauge sector that triggers top-quark condensation.  
Forbidding a large hierarchy in the
gap equation implies that top quark condensation is a spectator
to electroweak symmetry breaking, which must be accomplished 
mainly by another sector.  The properties of the 
electroweak
symmetry breaking scalars are identified.  Production mechanisms and
decay modes are studied. Unlike the standard model,
the scalar degree
of freedom most relevant to electroweak symmetry breaking 
can only be produced by its gauge interactions.  An
$e^\pm e^\pm \mu^\mp + \mu^\pm \mu^\pm e^\mp$ signal is proposed
to help unambiguously detect the presence of such a gauge-coupled
Higgs if it is light.  Other useful modes of detection are also
presented, and a summary is made of the search capabilities
at LEPII, Tevatron and LHC.

\end{abstract}

\end{titlepage}

\baselineskip=18pt
%\baselineskip=33pt

%\setcounter{footnote}{0}
%\setcounter{page}{1}
%\setcounter{figure}{0}
%\setcounter{table}{0}

%%%%%%%%%%%%%%%%%%%%%%%%%%%%%%%%%%%%%%%%%%%%%%%%%%%%%%%%%%%%%%%%%%%%%%%
%sss

\section{Finetuning and the top-quark gap equation}
\bigskip

As it became clear that the top quark is very massive
attention was naturally turned to top-quark condensate
models~\cite{nambu89,miransky89,bardeen90} which rely on
the gauged Nambu-Jona-Lasinio (NJL) model~\cite{nambu61}.  
Starting with a four fermion interaction
\beq
\label{njl}
L=L_{kin}+G\bar \psi_L t_R \bar t_R \psi_L
~~~\mbox{where}~~\psi_L=(t_L,b_L)
\eeq
a gap equation can be formulated for the dynamical top quark mass
\beq
\label{gapeq}
G^{-1}=\frac{N_c}{8\pi^2}\left( \Lambda^2-m_t^2\log \frac{\Lambda^2}{m_t^2}
      \right).
\eeq
Furthermore, in the fermion bubble approximation, with the above
gap equation enforced, zeros of the inverse
gauge boson propagators are found at non-zero momentum. This gives mass
to the gauge bosons and corresponds~\cite{pagels79,bardeen90,gherghetta94} 
to a decay constant of
\beq
\label{pagels}
{f^2_{\pi_t}}\simeq \frac{N_c}{16\pi^2}m_t^2\log 
        \frac{\Lambda^2}{m_t^2}.
\eeq
This relation is often referred to as the Pagels-Stokar formula.
In order for the top-quark condensate to account for all of electroweak
symmetry breaking, $f_{\pi_t}$ needs to be equal to $v=175\gev$.  
The currently measured top quark mass is $m_t=175\pm 6\gev$~\cite{top}, which
means that $\Lambda$ would have to be about thirteen orders of magnitude above
$m_W$
to account for all of electroweak symmetry breaking, an uninspiring
result for pure top-quark condensate models.

The Pagels-Stokar relation is logarithmically sensitive to the
condensate scale, and provides only a relationship between the
decay constant and dynamical quark mass, not the overall scale of
these parameters.  The gap equation, on the
other hand, sets the overall scale of $m_t$ and therefore $f_{\pi_t}$.
The quadratic sensitivity to the condensate scale induces a large hierarchy
problem for the weak scale ($m_t$ and $f_{\pi_t}$) if 
$\Lambda$ is above a few TeV.  For large $\Lambda$ the four-fermion
coupling $G$ must be tuned to one part in $\Lambda^2/m_t^2$.  For a
condensate scale in the multi-TeV region, this finetuning is greater
than one part in $10^3$.  It will be assumed here that finetunings much
above this are unnatural, and are probably not maintained by 
nature~\cite{clark91}.
Therefore, from the Pagels-Stokar relation the decay constant associated
with top-quark condensation is $f_{\pi_t}\lsim 60\gev$, implying
that top-quark condensation is a spectator to electroweak symmetry
breaking (EWSB).  That is, top-quark condensation mainly provides the top quark
its mass, but another sector mainly provides the $W$ and $Z$ bosons
their masses.

The above argument have led to the conception of top-quark condensate
assisted models such as those discussed in 
Refs.~\cite{hill94,lane95,kominis95,buchalla96}.
Much emphasis and study have been devoted to the boundstate scalars and
gauge bosons
associated with top-quark condensation in these 
%models~\cite{hill94,lane95,kominis95,buchalla96,others,burdman96}.  
models~\cite{hill94}-\cite{burdman96}.  
Attention is given here to the degrees of freedom which bear the most
burden for electroweak symmetry breaking.  In the following, a single
electroweak symmetry breaking doublet is postulated 
to give mass to the $W$, $Z$
and all light fermions, and perhaps a small contribution to the top
and bottom quarks.  The production and decay modes of this doublet
are detailed.  The results will be significantly different than the
standard model doublet, since unlike the standard model the resulting
scalar field has a small coupling to the top quark.  The complexity
of the electroweak symmetry breaking sector will have little effect
on the detectability of the scalars associated with electroweak symmetry
breaking.

%%%%%%%%%%%%%%%%%%%%%%%%%%%%%%%%%%%%%%%%%%%%%%%%%%%%%%%%%%%%%%%%%%%%%%%%
\section{Top-quark condensate scalars}
\bigskip

Top-quark condensation only requires a critical four-fermion coupling
in a lagrangian such as that given in Eq.~\ref{njl}.  
Explaining the origin of such a
strong interaction is a more difficult question.  Topcolor is one such
explanation~\cite{topcolor,hill94,lane95,buchalla96}, 
which is used as an example here.
Upon integrating out the heavy vector bosons from the topcolor
gauge group and the ``tilting'' $U(1)$ gauge group, one obtains
the following NJL interaction lagrangian:
\beq
L=L_{kin}+G_t\bar\psi_L t_R \bar t_R\psi_L
        +G_b\bar \psi_L b_R \bar b_R \psi_L.
\eeq
Tuning the couplings of the different gauge interactions which
formed this four fermion interaction lagrangian, one can obtain
$G_t>8\pi^2/N_c\Lambda^2>G_b$, which induces a $\langle t\bar t\rangle$
condensate but not a $\langle b\bar b\rangle$ condensate.

It is convenient to introduce auxiliary fields $\phi_t$ and $\phi_b$
and rewrite the above lagrangian as
\beq
L=L_{kin}-(\bar \psi_L\phi_t t_R + \bar \psi_L \phi_b b_R +h.c.)
       -G^{-1}_t \phi^\dagger_t\phi_t-G^{-1}_b \phi^\dagger_b \phi_b.
\eeq
Renormalization group evolution~\cite{bardeen90,buchalla96,blumhofer95} 
below $\Lambda$ makes the scalar
fields $\phi_t$ and $\phi_b$ dynamical, and it induces a vacuum
expectation value for $\phi_t$ (not $\phi_b$) in accord with the gap equation.
The pion decay constant of $\phi_t$ can be solved via the Pagels-Stokar
formula of Eq.~\ref{pagels} 
given particular values of $m_t$ and $\Lambda$.  For
the purposes of this paper we will assume the 
numerical value $f_{\pi_t} \simeq 50\gev$.

In the approximation that $f^2_{\phi_t}\ll v^2$ ($v$ is
normalized to $175\gev$)
top condensation is merely a spectator to electroweak symmetry
breaking (in analogy to chiral symmetry breaking among the light quarks)
and so they do not mix into the longitudinal components of the $W$
and $Z$.  Therefore, all the component states of $\phi_t$ and $\phi_b$
are physical eigenstates to a good approximation,
\beq
\phi_t = \left( \begin{array}{c} 
                  f_{\pi_t}+\frac{1}{\sqrt{2}}(\htz+ia^0_t) \\
                   h^\pm_t 
                \end{array} \right),~~~~~
\phi_b = \left( \begin{array}{c} 
                 h^\pm_b    \\
           \frac{1}{\sqrt{2}}(\hbz+ia^0_b) 
                \end{array} \right) .
\eeq
In the fermion bubble approximation the scalar $t\bar t\to t\bar t$ scattering
amplitude has a pole at $p^2=4m^2_t$ which can be identified with the mass
of the $\htz$.  Other masses are dependent upon details of the
underlying model and can vary from about $150$ to 
$350\gev$~\cite{hill94,kominis95,buchalla96}.
The top quark gets almost all its mass by a large Yukawa coupling to $\phi_t$,
and it can be assumed that the bottom quark gets almost all its mass 
from topcolor instanton effects~\cite{hill94,buchalla96}.

In addition to these boundstate topcolor scalars, there needs to
be degrees of freedom that break $SU(2)\times U(1)_Y$.  To do this
we introduce another doublet, $\phi_{ew}$, whose vacuum expectation
value is such that
\beq
f_{\pi_t}^2+f_{\pi_{ew}}^2=v^2=(175\gev)^2.
\eeq  
We also assume
that this doublet is a fundamental scalar and
gives a small mass to all the light 
fermions in the theory in addition to breaking
electroweak symmetry.  Even though the following assumes a fundamental
scalar breaking electroweak symmetry, 
it might be that $SU(2)\times U(1)_Y$ is
broken by some other strongly interacting sector which produces a narrow
resonance composite $\phi_{ew}$ which has very similar properties to the
fundamental Higgs state discussed below.

It will be convenient later to introduce a ``top-quark condensate 
spectator angle''
$\theta_t$, which is defined as
\bea
f_{\pi_{ew}} & = & v\sin\theta_t \\ 
f_{\pi_t} & = & v\cos\theta_t .
\eea
Using the numerical value of $f_{\pi_t}=50\gev$, one obtains
\beq
\cos\theta_t=\frac{50\gev}{175\gev}\simeq 0.3 \, .
\eeq
The pseudoscalar and charged
Higgs fields in the $\phi_{ew}$ doublet all get eaten by the $W^\pm$ and 
$Z$ bosons, and what remains physical is a single scalar field $\hewz$, whose
mass cannot be determined {\it a priori}.  Note also that since
$f_{\pi_t}\ll v$, and since $\hewz$ does not couple to tops (or couples very
weakly), then $\hewz$ is a mass eigenstate.

The assumption of just one scalar providing electroweak symmetry breaking
is not important for the qualitative results below.  The most
important assumption is that a $t\bar t$ condensate gives the top
quark its large mass.  If we added more condensing
scalar fields to the spectrum, it is possible to have large Yukawa couplings
of these scalars to the light quark fields.  
However, in
this case these additional fields 
would also have to be spectators to electroweak symmetry breaking,
just as $\phi_t$ is, and there would remain the necessity of a scalar(s) with
large vacuum expectation values which have small Yukawa interactions
with the fermions.

If the electroweak symmetry breaking
sector were more complicated, more angles would need to be introduced.  
For example, if a two Higgs doublet model
were involved then a new angle, $\beta$, would need to be introduced which
specifies the ratios of the vevs of these two fields.  Defining 
$\tan\beta \equiv \langle \phi_{ew,2}\rangle /\langle \phi_{ew,1}\rangle$,
then
\bea
f_{\pi_t} & = & v\sin\theta_t \\
f_{\pi_{ew,1}} & = & v\cos\theta_t\cos\beta \\
f_{\pi_{ew,2}} & = & v\cos\theta_t\sin\beta .
\eea
This angle $\beta$ does not enhance the coupling of the neutral
scalars to the top-quark
but does reduce the coupling to gauge bosons by a factor of
$\sin^2(\beta -\alpha)$ or $\cos^2(\beta -\alpha )$ where $\alpha$
is the mass eigenstate mixing angle between $h^0_{ew,1}$
and $h^0_{ew,2}$.  The argument generalizes to an arbitrarily 
complex electroweak symmetry breaking sector with doublets to show
that Yukawa production processes are negligibly affected by additional
complexity in the EWSB sector, and gauge production processes can
only be {\it reduced}. Thus, production modes for the scalars will decrease
with additional complexity.  Higher dimensional $SU(2)$ representations
such as triplets are not considered since they explicitly break 
custodial $SU(2)$ symmetry.  Barring finetuned cancellations, triplets
are incapable of providing a substantial fraction of EWSB due to the 
current $\rho$ parameter constraints~\cite{rizzo}.
Since our conclusion will be that the gauge-coupled
EWSB scalars are extremely difficult to reconstruct at the LHC, it is
most conservative to maximize the gauge boson couplings of the EWSB
scalar(s) by postulating
one single $\phi_{ew}$ which breaks
electroweak symmetry, and analyzing the consequences of this assumption.

%%%%%%%%%%%%%%%%%%%%%%%%%%%%%%%%%%%%%%%%%%%%%%%%%%%%%%%%%%%%%%%%%%%%%%%
\section{Decays of the $\hewz$}
\bigskip

The decay of the $\hewz$ (the neutral scalar component of $\phi_{ew}$) 
has a few inherent uncertainties.  The most important
uncertainty is how much of the $b$ quark mass is derived from $f_{\pi_{ew}}$.
This sets the coupling of the $\phi_{ew}$ to the $b$ quark, which in turn
sets the partial width of $\hewz\to b\bar b$.
The instanton induced $b$ quark mass~\cite{hill94,buchalla96} 
can be substantial and provide the
$b$ quark with all of its mass, leaving a very small (if any) residual
$b$ quark mass contribution from $\phi_{ew}$.  Furthermore, the hierarchy
of masses generated by $\phi_{ew}$ need not be in descending order of
generations.  That is, the $b$ quark mass induced by $\phi_{ew}$ could
be much smaller than the strange quark.  The working assumption will be
that the $b$ quark coupling to $\phi_{ew}$ is negligible.  However, it
cannot be ruled out that all of the $b$ quark mass comes from
$\phi_{ew}$ and this possibility will be commented on later.

In the standard model all the couplings to the gauge bosons and fermions
are set without ambiguity, and the decay branching fractions 
can be calculated~\cite{higgsreview,spira95,kunszt96}.
In Fig.~\ref{smdecay} the decay branching fractions of the standard model
are displayed for  $50\gev <m_{\hsmz}<200\gev$.  
%%%%%%%%%%%%%%%%%%%%%%%%%%%%%%%%%%%%%%%%%%%%%%%%%%%%%%%%%%%%%
\jfig{smdecay}{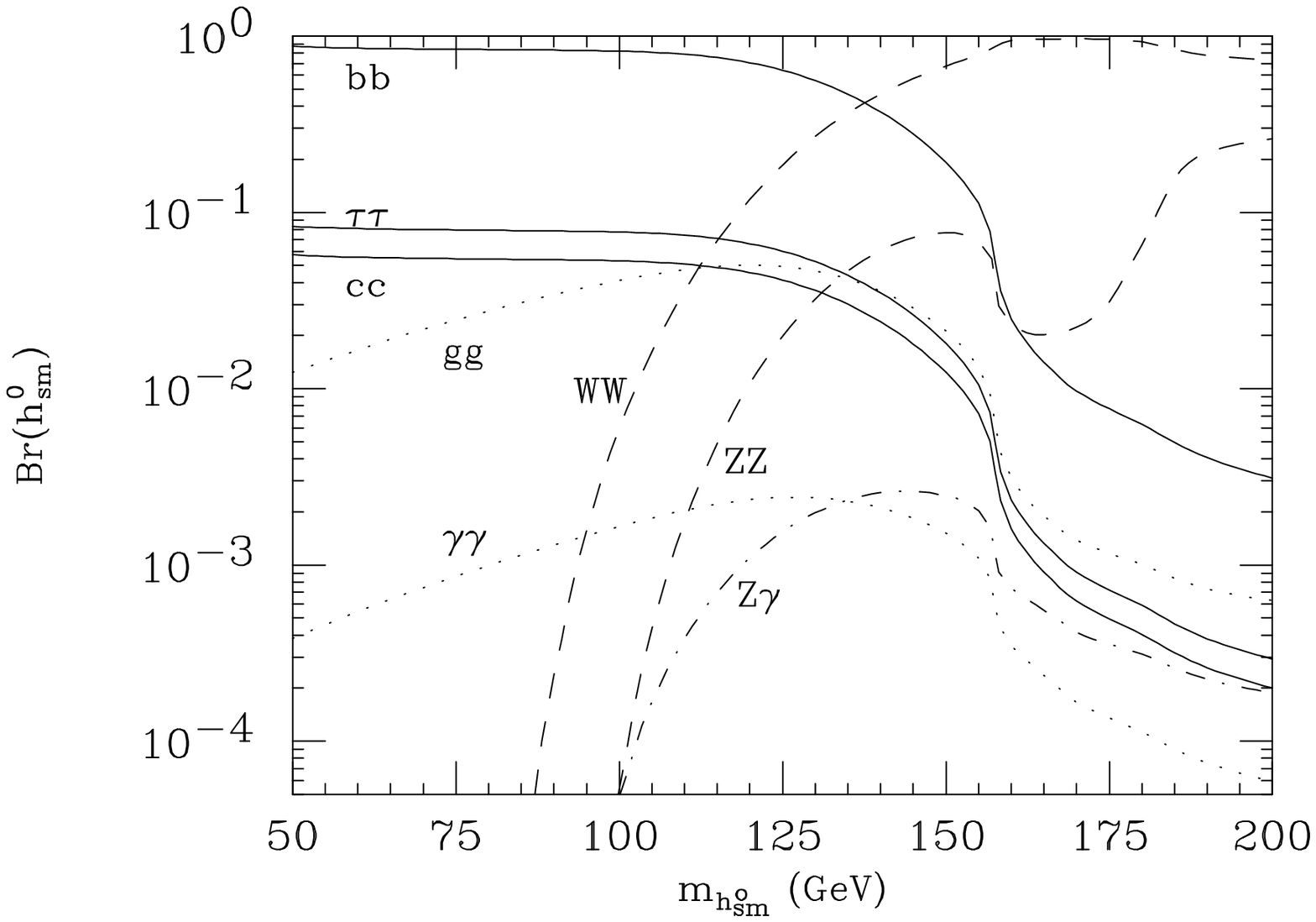}{Branching fractions of
${\hsmz}$ versus its mass.}
%%%%%%%%%%%%%%%%%%%%%%%%%%%%%%%%%%%%%%%%%%%%%%%%%%%%%%%%%%%%%
Above $200\gev$ the $WW$ and
$ZZ$ branching fractions dominate in a ratio of 2 to 1 for a Higgs mass well
above the $ZZ$ threshold.  Above the $t\bar t$
threshold the branching ratio to $t\bar t$ reaches its maximum of $20\%$ at
$m_{\hsmz}=500\gev$ and slowly reduces to $10\%$ at $m_{\hsmz}=1\tev$.
($m_t=175\gev$ is assumed throughout this discussion.)
Note the dominance of the $b\bar b$ final state up to $m_{\hsmz}\lsim 135\gev$.
Above this, the branching fraction into $WW$ becomes dominate.

The branching fractions of $\hewz$ can likewise be calculated once the
angle $\cos\theta_t$ is known, and the $b$ quark coupling to $\hewz$ is
specified.  In Fig.~\ref{ewdecay} the decay branching fractions are
shown for $\cos\theta_t=0.3$ and negligible $b$ quark coupling to $\hewz$.
%%%%%%%%%%%%%%%%%%%%%%%%%%%%%%%%%%%%%%%%%%%%%%%%%%%%%%%%%%%%%%%%%%%%%%%
\jfig{ewdecay}{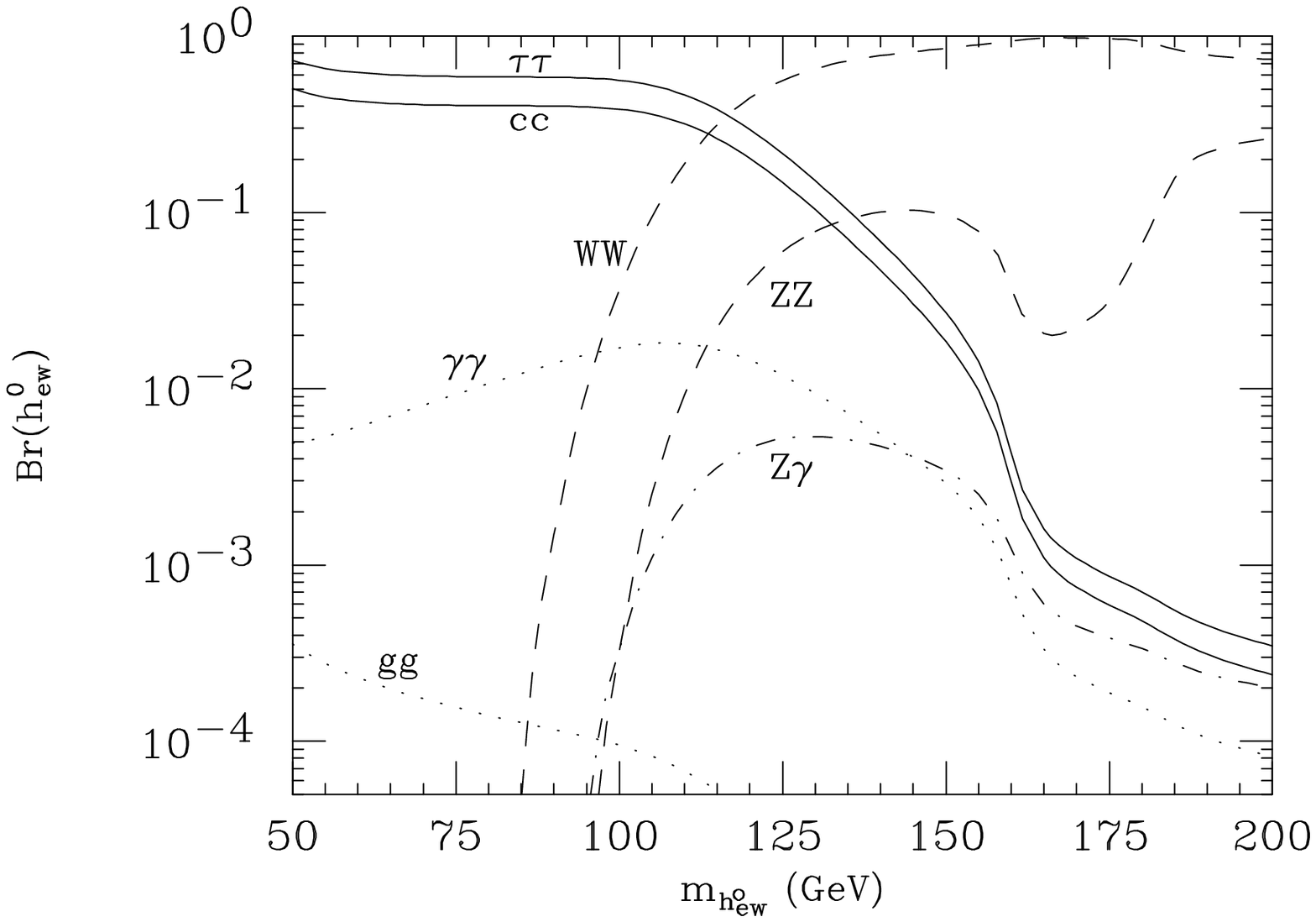}{Branching fractions of $\hewz$ versus
its mass for $\cos\theta_t=0.3$ and negligible $b$ quark coupling
to $\hewz$.}
%%%%%%%%%%%%%%%%%%%%%%%%%%%%%%%%%%%%%%%%%%%%%%%%%%%%%%%%%%%%%%%%%%
The $\tau^+\tau^-$ and $c\bar c$ final states are more important
for $\hewz$ than for $\hsmz$.
Furthermore the $WW^*$ branching fraction is much enhanced despite
the $\sin^2\theta_t$ suppression of the partial width to $WW^*$.
The branching fractions of $\hewz$ are much different than the standard
model Higgs for $m_{\hewz}\lsim 140\gev$.  Above $140\gev$ the $\hewz$
branching fractions are similar to the standard model except that
$\hewz\to t\bar t$ always has a negligible rate.  A further
important distinction from the standard model
is the substantially lower production
rate of $\hewz$ compared to $\hsmz$ at a hadron collider discussed in
the next section.

%%%%%%%%%%%%%%%%%%%%%%%%%%%%%%%%%%%%%%%%%%%%%%%%%%%%%%%%%%%%%%%%%%%%%%%%%%
\section{Production of $\hewz$}
\bigskip

The production modes of a standard model Higgs boson can be broken up
into three categories:
\begin{itemize}
\item Gauge processes: $q\bar q\to Z\hsmz$, $q\bar q'\to W^\pm\hsmz$,
                       $WW(ZZ)\to \hsmz$,
                       $e^+e^-\to Z\hsmz$, $e^+e^-\to e^+e^-\hsmz$.
\item Yukawa processes: $q\bar q\to t\bar t\hsmz$, $gg\to \hsmz$,
                        $e^+e^-\to t\bar t\hsmz$,
\item Gauge or Yukawa processes: $\gamma\gamma\to \hsmz$.
\end{itemize}
The $gg\to \hsmz$ production mechanism is called a ``Yukawa process'' because
it has a non-negligible rate due to the large top Yukawa coupling
of the $\hsmz$ to the top quark in the loop.  The $\gamma\gamma\to \hsmz$
is called a ``Gauge or Yukawa process'' since it has non-negligible 
contributions from both $W$ and top loops.

The scalars in top-quark condensate 
models~\cite{hill94,kominis95,buchalla96,burdman96}
generally either get produced by gauge processes
{\it or} by Yukawa process.  For example, the top Higgs, $\htz$, has a small
gauge production process suppressed relative to the standard model
by $\cos^2\theta_t\simeq 0.08$.  On the other hand, its production via
Yukawa processes is enormous due to the large Yukawa coupling of the
top quarks to $\phi_t$.  The bottom Higgs, $\hbz$, has approximately zero
production cross section via gauge processes, however, it has a reasonable 
Yukawa induced cross section from $gg\to \hbz$ and even 
$qq(e^+e^-)\to b\bar b\hbz$.

On the other hand, the electroweak symmetry breaking scalar, $\hewz$,
has negligible production cross section due to Yukawa processes.
It must be produced via gauge interactions, which are slightly suppressed
compared to the gauge interactions of the 
standard model Higgs boson.  The cross sections are related
to the standard model cross sections by
\beq
\sigma_{gauge}(\hewz) =\sin^2\theta_t \sigma_{gauge}(\hsmz)
\eeq
where $\sigma_{gauge}(h)$ is either $\sigma(Wh)$, $\sigma(Zh)$
or $\sigma(WW(ZZ)\to h)$, etc.
For $f_{\pi_t}=50\gev$ the suppression factor is 
merely $\sin^2\theta_t\gsim 0.92$.

The lack of Yukawa production processes for $\hewz$ is a severe
handicap for a search at the LHC. The dominant production cross section
for the standard model Higgs is $gg\to \hsmz$.  
In Fig.~\ref{hprod} the cross sections for $jj\hewz$, $W\hewz$,
and $Z\hewz$ at the LHC ($\sqrt{s}=14\tev$) are shown.  The $jj\hewz$
cross section is from $WW(ZZ)\to \hewz$ where the two jets are the
ones which radiate the $W$'s and are generally at high rapidity.
%%%%%%%%%%%%%%%%%%%%%%%%%%%%%%%%%%%%%%%%%%%%%%%%%%%%%%%%%%%%%%%%%%%5
\sfig{hprod}{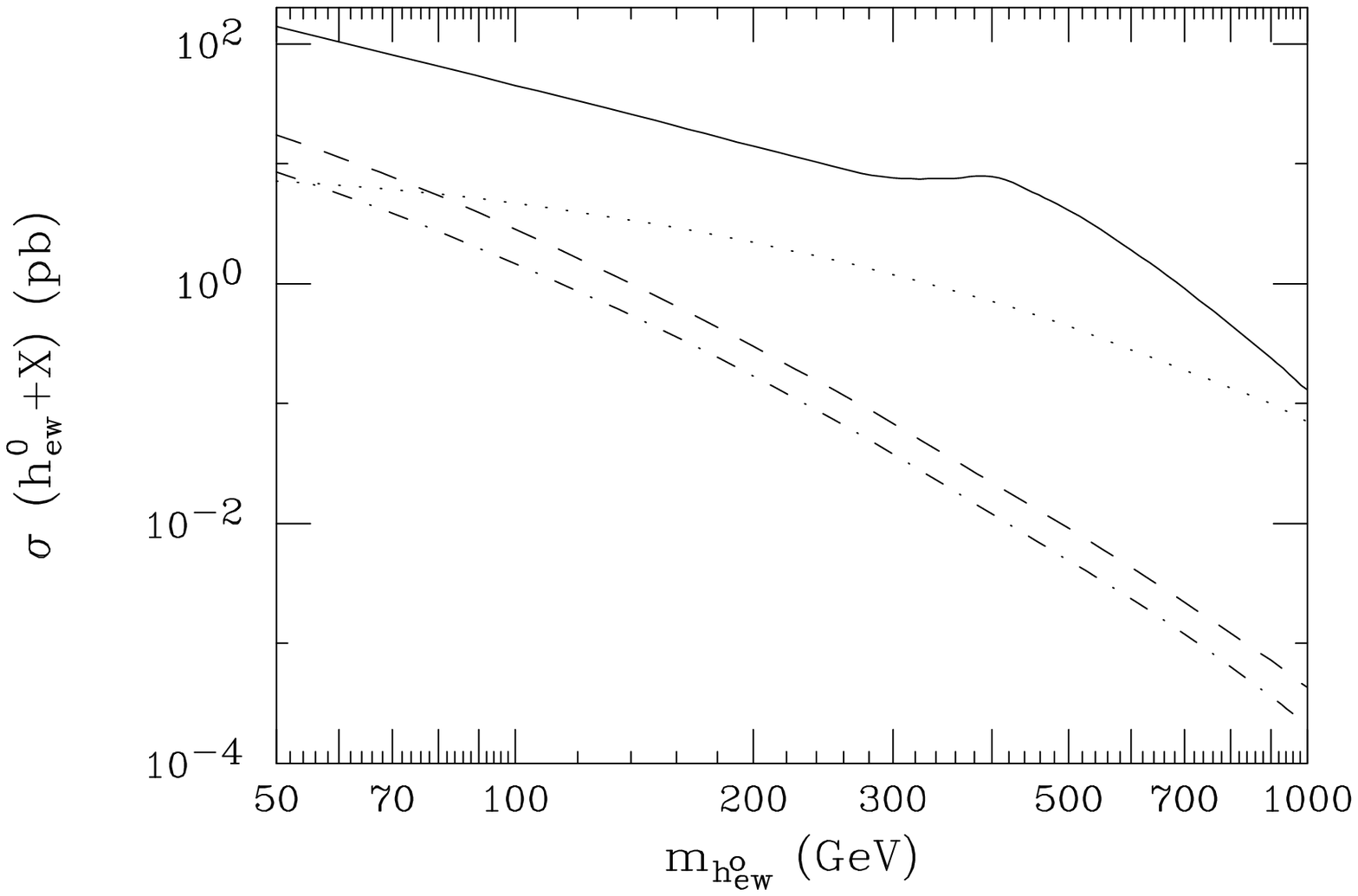}{Cross sections for $jj\hewz$ (dotted), 
$W\hewz$ (dashed),
and $Z\hewz$ (dot-dashed) 
at the LHC ($14\tev$) for $\cos\theta_t =0.3$.  Also, the standard model Yukawa
production cross section $gg\to \hsmz$ (solid)
is shown to demonstrate
the relatively low instrinsic rate of the gauge production cross sections.}
%%%%%%%%%%%%%%%%%%%%%%%%%%%%%%%%%%%%%%%%%%%%%%%%%%%%%%%%%%%%%%
We see that the total production rate for $\hewz$ is much
smaller than for $\hsmz$ due to lack of the $gg\to \hewz$ Yukawa mode.
The $\hewz\to \gamma\gamma$ signal for $100\gev\lsim m_{\hewz}\lsim 140\gev$
is no longer viable since the total rate of Higgs production does
not overcome the small (albeit enhanced) branching fraction to
two photons.  Detecting $\hewz$, if it is possible, 
must rely on the smaller gauge
production cross sections.

%%%%%%%%%%%%%%%%%%%%%%%%%%%%%%%%%%%%%%%%%%%%%%%%%%%%%%%%%%%%%%%%%%%%%%%
\section{Detecting the $h^0_{ew}$}
\bigskip

The search for the standard
model Higgs boson has been well studied and the strategy is well 
mapped~\cite{higgsreview}.  LEPII running at $\sqrt{s}=192\gev$ will
be able to see the Higgs boson up to about $95\gev$ with $150\,\mbox{pb}^{-1}$
per experiment.  
Above this, the LHC experiments should be able to see the
standard model Higgs through $gg\to \hsmz\to \gamma\gamma$ up to
about $140\gev$, and through $gg\to \hsmz\to ZZ^{(*)}$ if 
$140\gev \lsim m_{\hsmz}\lsim 700\gev$.  For scalar resonances above
about $700\gev$ the concept of a 
perturbatively coupled Higgs boson becomes more
tenuous, and longitudinal $W$ scattering analyses become more important.

The common thread among all the searches described above is the
$gg\to \hsmz$ Yukawa production mode.  In the standard model this
process dominates the production rate for Higgs bosons.  Some studies
have been performed on the ability to detect the Higgs boson via
$W\hsmz$~\cite{stange94,mrenna94,was95,agrawal95}, which is a gauge-coupled
production mode.  The hope is that with
two tagged $b$ quarks from $\hsmz$ and a high $p_T$ lepton tag
from the $W$, the small rate of $W\hsmz$ could still be pulled out
from the background ($WZ$, $Wb\bar b$, $t\bar t$, etc.).  At the LHC
it was concluded by~\cite{was95} that this mode could be useful if
$80\gev\lsim m_{\hsmz}\lsim 100\gev$, whereas~\cite{agrawal95} finds
that $m_{\hsmz}\lsim 130\gev$ are detectable in this mode
with excellent bottom jet identification.

For the $\hewz$ state, the LEPII capability should not be altered.
The $\tau$ branching fraction of the Higgs will be large and the
backgrounds from $ZZ$ where one $Z$ decays to $\tau^+\tau^-$ will
be relatively small since that branching fraction is only about
$3\%$. Searches at the LHC with $m_{\hewz}\gsim 100\gev$ are
more difficult.
In the range from $100\gev \lsim m_{\hsmz}\lsim 135\gev$ the branching
ratio to $\tau^+\tau^-$ is greater than $8\%$.  The signal
$pp\to W\hsmz$ where $W\to jj$ and $\hsmz\to \tau^+\tau^-$ has been
studied in Ref.~\cite{mrenna94}.  It was concluded that a significant
signal at the LHC could not be found in this mode for a standard model
Higgs boson.  However, such a signal might be possible in this
$\tau^+\tau^-$ decay mode at the Tevatron running
at $\sqrt{s}=2\tev$ with $30\, \mbox{fb}^{-1}$ of integrated luminosity.
Turning back to the $\hewz$, the $\tau^+\tau^-$ decay branching fraction
is much increased.  The significance of the 
standard model $\tau^-\tau^+$ signal now
should be multiplied by a factor of 
$Br(\hewz\to\tau^+\tau^-)/Br(\hsmz\to \tau^+\tau^-)\simeq 6$.  It therefore
might be possible to detect the light $\hewz$ 
at the Tevatron with compelling significance if
$m_{\hewz}\lsim 130\gev$.  Furthermore, this signal might also be detectable
at the LHC with sufficient luminosity.

Earlier we postulated that $\hewz$ does not couple to the $b$ quarks.
If we now relax that assumption and assume the opposite extreme, that
all the $b$ quark mass comes from $\phi_{ew}$, then the branching fraction
of $\hewz$ into fermions for $m_{\hewz}\lsim 130\gev$ will be almost
identical to the standard model.  Then the analyses of the $W\hewz$
signal will become most important, and from studies in the standard 
model~(especially~\cite{agrawal95})
it can be concluded that a $b$ quark coupled $\hewz$ is detectable at
the LHC with as low as $10~\mbox{fb}^{-1}$ of data.  The Tevatron probably
could not detect this $b$-coupled Higgs state above the LEPII limit of 
$95\gev$~\cite{stange94,was95}.

If $m_{\hewz}\gsim 130\gev$ then it appears that the Tevatron and 
LHC will have 
difficulty seeing a signal and reconstructing the $\hewz$ mass.
In this range $\hewz$ will decay to $WW^*$ most of the time.  Since the
production rate for $\hewz$ is small the $ZZ^{(*)}$ decay cannot
be separated from backgrounds to construct a signal.  Without this
mode at our disposal it appears near impossible to measure the Higgs
mass since the $WW^{(*)}$ decay mode does not allow the Higgs mass
to be reconstructed.

However, it might be possible to detect a Higgs signal without being
able to reconstruct the mass.  If the Higgs decays to $W^+W^-$ (either
on-shell or off-shell) then the $pp\to W^\pm \hewz$ produces
a three lepton plus $\slashchar{E_T}$ signal if all the $W$'s decay
leptonically.  This signal is relatively clean in a hadron collider
environment and has been proposed as a search tool
for supersymmetric gaugino production of $\widetilde W\widetilde Z$
with subsequent leptonic decays into three leptons plus missing energy
carried away by the stable lightest supersymmetric particle~\cite{susy}.  
Because of this potentially large supersymmetric rate, 
it may be
difficult to tell the difference between a $W\hewz\to lll+\slashchar{E_T}$
signature as opposed to a $\widetilde W\widetilde Z\to lll+\slashchar{E_T}$
signature.  

Some permutations of these trilepton decays yield a like-sign
dilepton signal with missing energy and little
background~\cite{stange94}.  Potential
backgrounds to like-sign dileptons include
$WZ$ production where one lepton comes from the $W$ and one lepton
comes from the $Z$.  Cutting out $m_{l^+l^-}\simeq m_Z$ reduces this
background. Another background comes from $t\bar t$ production where one
lepton comes from a $W$ decay in the $t$ decay and the like-sign lepton comes
from a $b$ decay in the $\bar t$ decay.  This background can be greatly
reduced by an effective jet veto of the hadronic activity of the $b$ quark,
requiring no tagged $b$ quark in the event,
and also requiring that all the leptons be isolated. 
Here, we propose a more discerning 
signal of like-sign dileptons with a different flavor third
lepton.  
Using just the first two generations of leptons, the signal
for the presence of a Higgs boson is $e^\pm e^\pm \mu^\mp$ and
$\mu^\pm \mu^\pm e^\mp$, with all leptons isolated.  
This is a useful signal to distinguish $W\hewz$ production as opposed
to the supersymmetric $\widetilde W\widetilde Z$ production since
like-sign dileptons from the latter require all three leptons to
be of the same flavor.
The branching fraction into
$WW^{(*)}$ is given in Fig.~\ref{ewdecay}.  Using the production cross
section from Fig.~\ref{hprod}
the total number of signal events in $100~\mbox{fb}^{-1}$ of data is
displayed in Fig.~\ref{ewdetect}.  
%%%%%%%%%%%%%%%%%%%%%%%%%%%%%%%%%%%%%%%%%%%%%%%%%%%%%%%%%%%%%%%%%%%%%%%
\sfig{ewdetect}{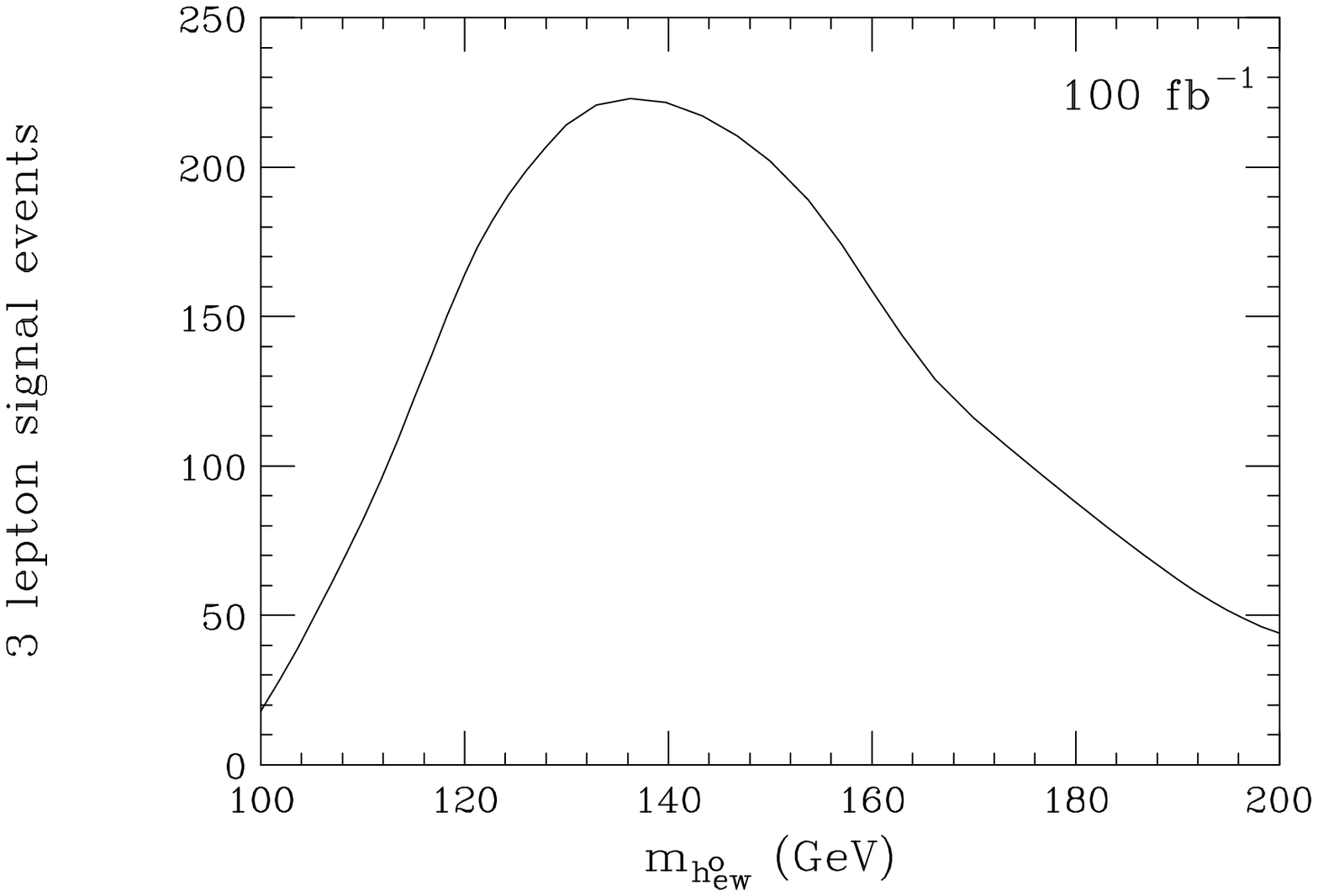}{The total event rate of $e^\pm e^\pm \mu^\mp+
\mu^\pm \mu^\pm e^\mp$ expected from $pp\to W\hewz\to WWW^{(*)}$ production
with $100\, \mbox{fb}^{-1}$ of data collected at the LHC with center of
mass energy $14\tev$. A $5\sigma$ signal above background requires
at least 100 signal events.}
%%%%%%%%%%%%%%%%%%%%%%%%%%%%%%%%%%%%%%%%%%%%%%%%%%%%%%%%%%%%%%%%%%%%%%

The main background from $t\bar t$ can be estimating using the geometric
acceptance of leptons ($|\eta |<2.5$) and jets ($|\eta |<5$), 
the combinatorics of the decays which
yield $e^\pm e^\pm \mu^\mp$ and
$\mu^\pm \mu^\pm e^\mp$ signal events, and the total inclusive
$t\bar t$ cross section at LHC, which will be conservatively bounded
at $2$~nb.  The number of background events estimated from $t\bar t$
is approximately $100$ for $100~\mbox{fb}^{-1}$ of data.  
(Note that there are no background events
from $WZ$.)  After applying the same lepton acceptance cuts
to the signal, a $5\sigma$ effect
would require at least 100 signal events (using
$S/\sqrt{B}$ for significance).  Thus, the electroweak symmetry
breaking Higgs, $\hewz$, could be detected in the range
$110\gev \lsim m_{\hewz}\lsim 175\gev$ by this process.  This estimate is
based solely on parton level calculations and is only meant to demonstrate
that this signal could be useful at the LHC.  Estimates of non-physics
backgrounds such as mis-identification, and full jet-level Monte Carlo
simulation could diminish the significance found here.
However, it is likely that additional cuts not studied here 
could further distinguish the $W\hewz\to WWW$
topology over $t\bar t\to WWbb$, thereby increasing the significance of the
signal.  Also, using the $\tau$ final states could help somewhat
since $b$ decays
to $\tau$'s are kinematically suppressed.

From Fig.~\ref{hprod}
we can see that the $WW(ZZ)\to \hewz$ vector boson fusion provides
the highest production rate for $\hewz$.  Along with the Higgs boson
comes two high rapidity jets corresponding to the quark lines which
radiated the initial vector bosons.  For low mass Higgses, where
$\hewz\to ZZ^{(*)}$, the signal would be four leptons with high
invariant masses, $m_{l^+l^-}$ from each decay.  The standard model
background is
dominated by continuum $Z\gamma^*$ production with enhancement at
small invariant $m_{l^+l^-}$.  Identifying the $Z^{(*)}$ as the lower
mass $m_{l^+l^-}$ and placing a cut on how low this invariant mass
very effectively removes the background while retaining almost all
of the signal~\cite{higgsreview}.  It might be possible to detect
the light $\hewz$ from vector boson fusion with as low
as $40$ signal events (before cuts) in this mode~\cite{higgsreview}.
The signal rate
for $VV\to \hewz \to ZZ^{(*)}\to l^+l^-l^{'+}l^{'-}$ ($l,l'=e~\mbox{or}~\mu$)
is shown in Fig.~\ref{zzrate}.
%%%%%%%%%%%%%%%%%%%%%%%%%%%%%%%%%%%%%%%%%%%%%%%%%%%%%%%%%%%%%%%%%%%%%%
\sfig{zzrate}{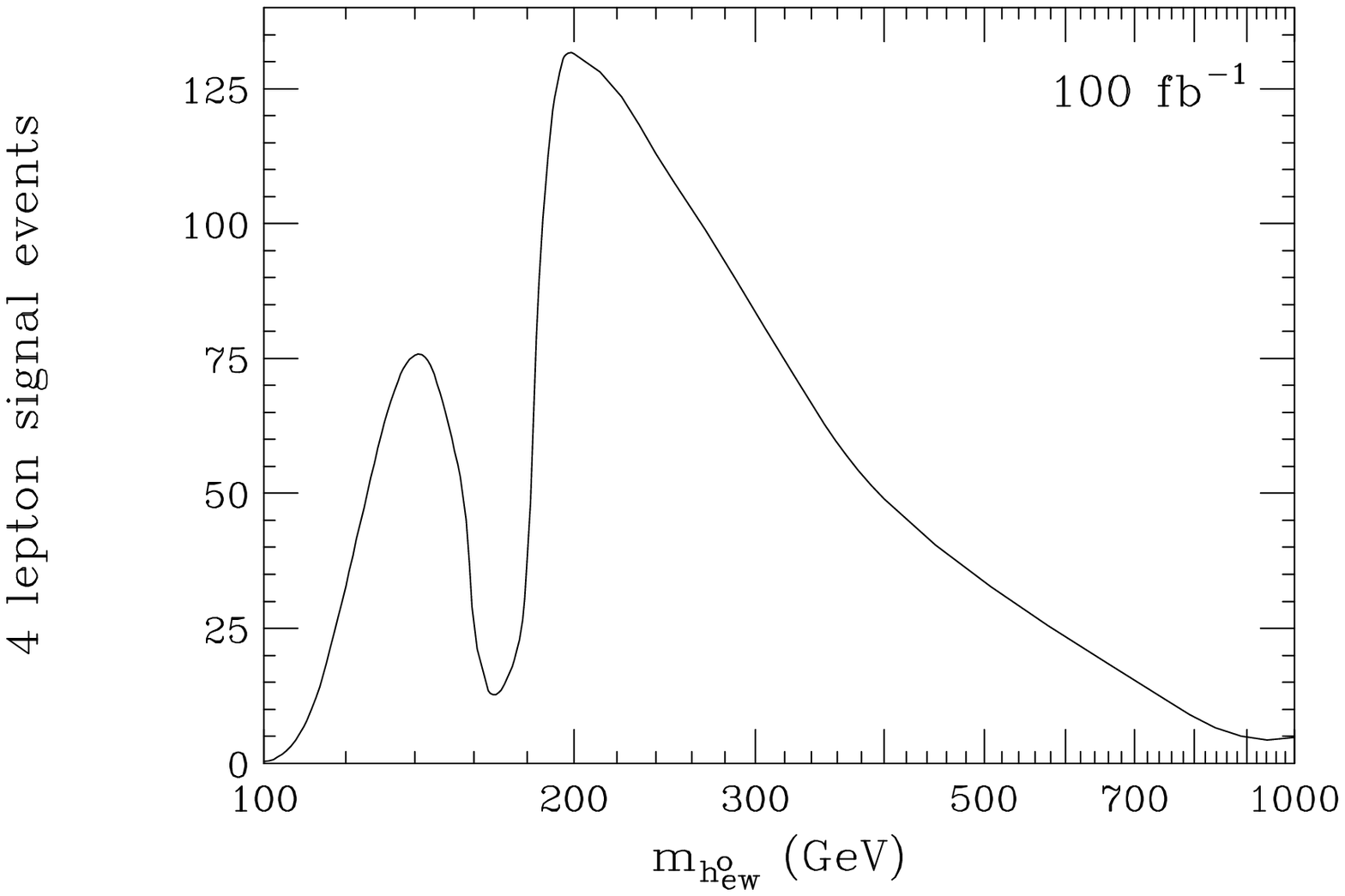}{The total event rate of 
$e^+e^-\mu^+\mu^-$ expected from $VV\to \hewz\to ZZ^{(*)}$ production
with $100\, \mbox{fb}^{-1}$ of data collected at the LHC with center of
mass energy $14\tev$. A $5\sigma$ signal above background from this
mode is estimated 
to be possible for the mass ranges $120\gev\lsim m_{\hewz}\lsim 155\gev$
and $200\gev\lsim m_{\hewz}\lsim 400\gev$.}
%%%%%%%%%%%%%%%%%%%%%%%%%%%%%%%%%%%%%%%%%%%%%%%%%%%%%%%%%%%%%%%%%%%%%%
In the $ZZ^*$ region, the detectability requirement of $40$ signal events
corresponds to a Higgs mass range of $120\gev\lsim m_{\hewz}\lsim 155\gev$.

The region between the $WW$ and $ZZ$ threshold is much more difficult.
The $\hewz\to ZZ^*$ branching fraction is greatly reduced in this region
because $\hewz\to WW$ is above threshold and rapidly increasing.  Therefore,
$VV\to \hewz\to ZZ^*$ is not viable.  From above it appears that
it might be possible to see $pp\to W\hewz\to WWW$ up to about $175\gev$.
However, above $175\gev$ the trilepton signal runs out. 
Thus, the small region between $175\gev \lsim m_{\hewz}\lsim 180$ must
rely on the difficult task of seeing a resonant enhancement in
$VV\to \hewz\to WW$.  
In the standard model, the large $gg\to \hsmz$ rate allows for the
possibility of seeing the Higgs in the $\hsmz\to WW$ channel for the
range $155\gev\lsim m_{\hsmz}\lsim 180\gev$~\cite{hww,dittmar96}.  
It has been estimated that with $5\, \mbox{fb}^{-1}$ a signal with
greater than $5\sigma$ significance could be observed in this
mass range for the standard model Higgs using the $\hsmz\to WW\to l\nu l'\nu'$
decay mode~\cite{dittmar96}.  The gauge-coupled Higgs $\hewz$ does
not get produced by $gg$ fusion but rather $WW(ZZ)$ fusion.  In this
mass range the total cross section is more than a factor of six lower
than the $gg$ cross section.  Taking this into account and scaling
the significance result of~\cite{dittmar96} up to $100\, \mbox{fb}^{-1}$
there is an $8\sigma$ signal available in the range
$155\gev\lsim m_{\hsmz}\lsim 180\gev$. It should be pointed out, however,
that some of the cuts employed to reduce background 
may not be as effective 
if the underlying signal
event is produced mainly by $WW(ZZ)$ fusion rather than $gg$ fusion.  
For example, the $|\cos\theta_{l^+l^-}|<0.8$ cut to reduce
continuum $q\bar q\to W^+W^-$ background may not be as useful.  However,
it might be possible to gain significance by tagging the high $p_T$,
high rapidity jets which radiate the initial vector bosons~\cite{cahn87}.
A central mini-jet veto~\cite{barger94}
which takes advantage of the different color flow of the signal compared to
background may also be helpful.  Again, confusion with a supersymmetric
signal is possible, especially for slepton and chargino pair production
which both can yield $ll'+\slashchar{E_T}$ signatures at high rates.  The
spectator jet tags would be especially useful to
eradicate this potential confusion.

Above the $\hewz\to ZZ$ threshold the Higgs signal would be a Higgs bump
in the $ZZ$ invariant mass spectrum in the total $ZZ$ production cross 
section.  Again, the $Z$'s are required to decay into leptons.  
Fig.~\ref{zzrate} give the total rate above the $ZZ$ threshold and can
be compared with the total $ZZ$ production cross section for sensitivity
of this invariant mass peak~\cite{ohnemus}.  The resonant Higgs signal
is smeared according to the detector's electromagnetic resolution,
which is taken to be $15\%/\sqrt{E_T}$.  Taking into consideration lepton
acceptance, the significance is calculated for signal events in a bin
centered on $m_{ZZ}=m_{\hewz}$ with a width determined
by the two standard deviation smearing of the Higgs resonance corresponding
to the resolution factor given above.  The significance reaches a peak of
about $7\sigma$ at $m_{\hewz}=220\gev$ and falls below $5\sigma$
for $m_{\hewz}\gsim 400\gev$.  This estimate is based only on the
$\hewz\to ZZ\to e^+e^-\mu^+\mu^-$ mode.  Using all lepton permutations
at the expense of potential lepton pairing ambiguity should extend the
search range well above $400\gev$.  
Again, a realistic detector simulation is required to obtain a precise range.
Nevertheless, it appears possible to discern a
Higgs peak up to high mass scales with $100\, \mbox{fb}^{-1}$.

In short, top-quark condensate models may imply at least one scalar degree
of freedom, $\hewz$, which 
breaks electroweak symmetry and has negligible Yukawa
production modes.  The LEPII search range should not be different than
the standard model.  The Tevatron $\hewz$ search is even enhanced over
the standard model Higgs search range.  On the other hand,
it will be more challenging at the LHC to 
demonstrate that such a gauge-coupled Higgs like $\hewz$ can be
detected, and most especially that its mass can be reconstructed.
However, over most of the perturbative
Higgs mass range detection
should be possible with $100\, \mbox{fb}^{-1}$ of data.

%%%%%%%%%%%%%%%%%%%%%%%%%%%%%%%%%%%%%%%%%%%%%%%%%%%%%%%%%%%%%%%%%%%%%%%%

\bigskip
\noindent
{\it Acknowledgements.} The author has benefited from 
discussions with G.~Buchalla, S.~Martin and M.~Peskin.

%%%%%%%%%%%%%%%%%%%%%%%%%%%%%%%%%%%%%%%%%%%%%%%%%%%%%%%%%%%%%%%%%%%%%%%%

%%%%%%%%%%%%%%%%%%%%%%%%%%%%%%%%%%%%%%%%%%%%%%%%%%%%%%%%%%%%%%%%%%%%%%%%

\end{document}